\documentclass[aps,prl,twocolumn,showpacs,superscriptaddress,groupedaddress]{revtex4}  
\usepackage{graphicx}  
\usepackage{dcolumn}   
\usepackage{bm}        
\usepackage{amssymb}   

\begin{document}

\widetext


\title{Mechanically controlled quantum switch defined on a curved 2DEG}

\author {S.Seyyare Aksu}
\author {Oguzhan Kasikci}
\author {Afif Siddiki}
\affiliation{Physics
Department, Faculty of Sciences and Letters, Mimar Sinan Fine Arts University, 34380-Sisli, Istanbul,
Turkey}

\date{\today}

\begin{abstract}
To investigate quantum nature of two dimensional electrons subject to high perpendicular magnetic fields, usually a planar electronic Fabry-P\'erot interferometer is utilized. In this work, we investigate an interferometer defined on a curved heterostructure. In the presence of a magnetic field perpendicular to the cylindrical axis, the location and the properties of the edge channels depend on the radial component of the magnetic field. Considering a curved structure, we perform numerical and semi-analytical calculations to determine widths of the incompressible edge states. We observe that the edge states form a closed loop for certain magnetic field strengths yielding observation of conductance oscillations, which can be manipulated by changing the Azimuthal angle mechanically. In addition, we investigate the effect of spin polarization on the edge state distribution considering Zeeman splitting and obtained odd integer edge states. The proposed experiment would yield a novel method to clarify the ongoing debate on the origin of conductance oscillations, namely whether they stem from Aharonov-Bohm phase or charging effects.
\end{abstract}

\maketitle

The quantum Hall based particle interferometers attracted significant attention in recent years, due to unexpected behavior of the conduction oscillations observed as a function of magnetic field.~\cite{PhysRevB.79.241304, Ofek23032010, PhysRevLett.98.106801} In experiments, a two path particle interferometer is defined by quantum point contacts (QPCs) acting as an optical beam splitter, meanwhile the beam is replaced by edge states resulting from the electrostatic confinement and the quantizing magnetic field. The interference pattern reflects itself as oscillations in the potential difference measured between the source and drain contacts, where drain contacts are essentially detectors. First, it is observed that the potential difference (or conductance) oscillates with certain periods depending on the magnetic field. Second, different from the optical interferometers, the Coulomb interaction between electrons becomes important in determining the period of the conductance oscillations. Interestingly, the magnetic field period $\Delta B$ exhibits two distinct behavior depending on the area of the interference loop, namely the sample size: For large samples ($\sim18 \mu$m$^2$), $\Delta B$ is constant indicating Aharonov-Bohm (AB) interference, whereas for small samples ($\sim2 \mu$m$^2$) period is linear in $B$, pointing Coulomb Blockade (CB). A second experimental parameter to change the loop area is to utilize an additional metallic gate to deplete electrons from the edges. Confusingly, the period of the oscillations in gate potential $\Delta V_g$ presents opposite behavior.
\begin{figure}
\scalebox{0.25}{\includegraphics{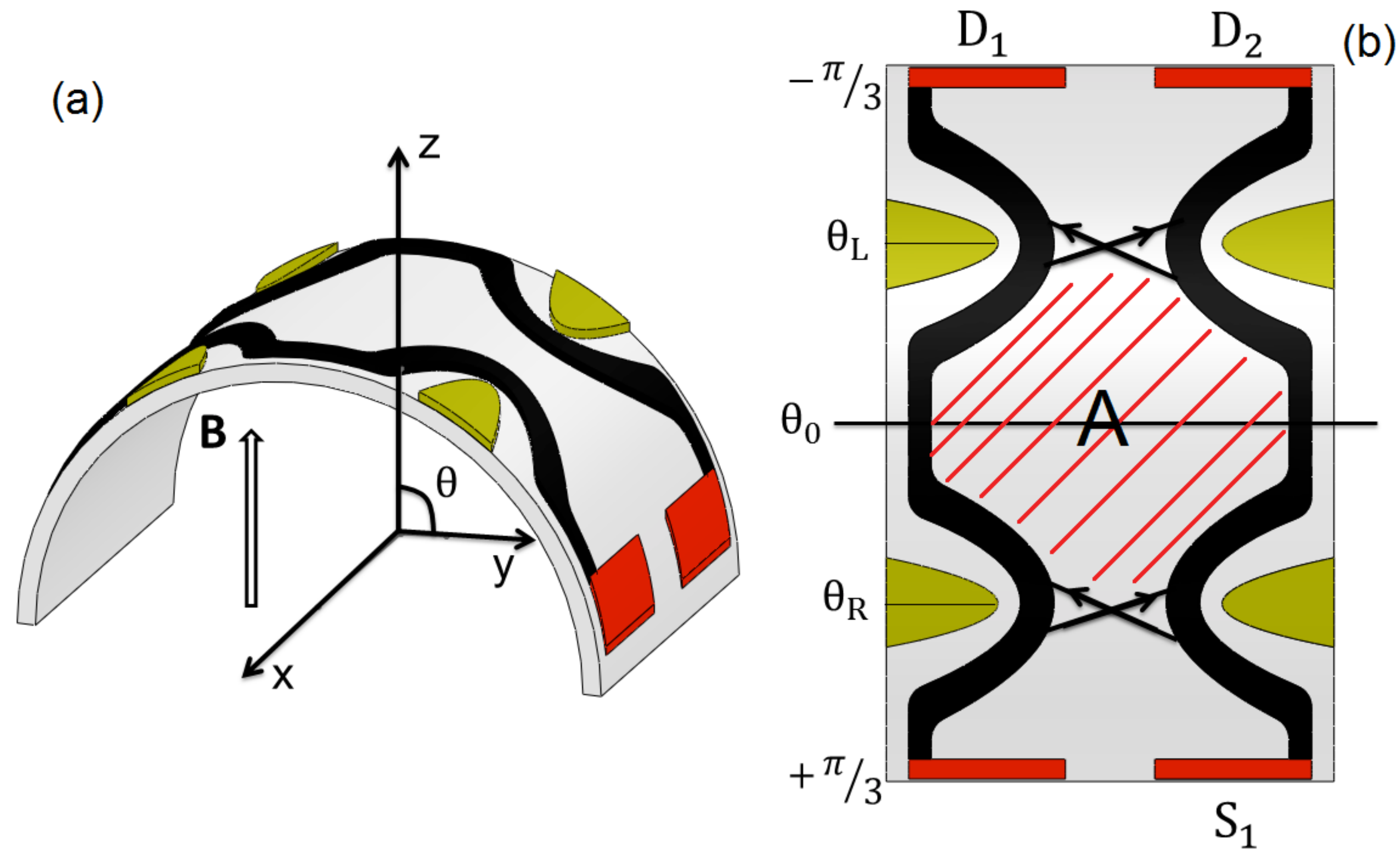}}
\caption{(a) 3D illustration of the curved interferometer. Light (yellow) regions depict QPCs, whereas source ($S_1$) and drain contacts ($D_{1,2}$) are shown by darker boxes. Incompressible strips are denoted by black. (b) The surface projection of the dot. The area $A$ of the closed loop formed by incompressible strips is patterned. Arrows show where the scattering (partitioning) takes place.}\label{fig1}
\vspace{-1 cm}
\end{figure}
At small samples $\Delta V_g$ is constant for all $B$ fields, whereas at the large samples $\Delta V_g$ varies linearly with inverse magnetic field. Within the main stream theories these observations are attributed either to AB phase of the particles or to charging effects. It is stated that, at small samples CB dominates $\Delta B$ whereas at large samples the it is due to AB phase of electrons~\cite{PhysRevB.79.241304}. However, there are other scenarios which rely on self-consistent calculations including time-dependent solution of the Schr\"odingier equation and claim that charging effects are not essential to explain the observations. Attempts to answer the question whether if the oscillations stem from AB or CB have yielded wildly divergent theories~\cite{1367-2630-14-5-053024, Cicek20101095}. From application point of view, these electronic interferometers can be utilized as a (quantum) switch, where the maxima of the oscillation corresponds to an open circuit and the minima mimicking a closed circuit controlled by the quantum mechanical properties of the system, for instance by the AB phase. Hence, it is important to understand the mechanism behind the oscillations. 

In this letter, we propose an electronic interferometer that can be defined on a curved heterostructure and calculate the positions of the edge states within the electrostatic approximation. Next, we obtain the widths of the incompressible strips (ISs) depending on the magnetic field strength and azimuthal angle of the curvature. This investigation allows us to calculate the number of magnetic flux quanta enclosed by the interference loop as a function of $B$, which presents a non-linear behavior. At the final step, we include the effect of spin polarization and investigate the areal dependency considering Zeeman split quantized levels.    

\textbf{AB interference conditions}

Experimentally, interferometers are fabricated on high-mobility GaAs/AlGaAs heterostrucutes, comprising a two dimensional electron system (2DES) at the interface. The interferometer is usually defined by metallic surface gates, providing physical edges, and two QPCs are utilized as constrictions. In the presence of a perpendicular magnetic field, the energy dispersion is (Landau) quantized and edge channels are formed, which spatially follow the confinement. Namely, the edge channels are the equipotential lines at the Fermi energy which are shifted in energy by Landau quantization. Utilizing the edge channels, a closed interference path can be formed to observe quantum interference where the electrons can tunnel from one edge channel to the other. The Aharonov-Bohm phase is measured when an electron travels along a closed path, where a magnetic field penetrates the loop.  The phase difference between two paths equals to $\Delta\phi=2\pi\Phi/\Phi_{0}$, $\Phi$ being the total magnetic flux and $\Phi_{0}=h/e$ is the magnetic flux quantum. Taking into account Coulomb interaction modifies the naive single particle picture by yielding formation of compressible and ISs. Across the compressible strip, the Fermi energy equals to Landau level and the system behaves like a metal due to the existence of the available states. Thus the electrons are redistributed where the total potential is approximately constant since screening is almost perfect. On the other hand, across the IS the Fermi energy lies between two successive Landau levels and the system behaves like an insulator due to vanishing density of states. Hence, electrons cannot be redistributed and the potential presents a variation, i.e. screening is poor. The occupation of the Landau levels are depicted by the filling fraction (or factor) $\nu$, which is the ratio between the number of electrons $N_{\rm el}$ and the number of flux quanta $N_{\rm \Phi_0}$. At an IS this ratio is an integer. As shown by Shklovskii and Fogler~\cite{1994PhRvB..50.1656F}, the external non-equilibrium current flows from the ISs due to suppressed scattering. Later, it was shown self-consistently that the current flows from the IS only if the width of the strip is larger than the thermodynamical (Fermi wavelength, $\lambda_F=\sqrt{2\pi/n_{0}}$) and quantum mechanical (magnetic length, $\ell_{B}=\sqrt{\hbar/eB}$) length scales~\cite{PhysRevB.70.195335}. Note that, incompressibility is a thermodynamical quantity, therefore, once the width of the strip becomes small or comparable to $\lambda_F$, statistically it is possible to have scattering along the ISs. However, if the strip width is larger than $\ell_{B}$ and narrower than $\lambda_F$ the strip is called an \emph{evanescent} incompressible strip and its existence is shown by experimental investigations~\cite{2013NatSR...3E3133K}. As a consequence of the thermodynamical properties of the ISs, the conductance oscillations can only be observed under the certain conditions which depends on the width of the IS and their mutual distance~\cite{1367-2630-14-5-053024, Cicek20101095}. The conditions to observe an interference pattern are to have partitioning (scattering) between two (evanescent) ISs and to get strips in close vicinity ($\sim\ell_B$) to prevent phase decoherence. 

At a planar 2DES the consequences of Coulomb interaction is discussed in the pioneering work of Chklovskii et al.~\cite{PhysRevB.46.4026}, considering spinless electrons. The IS (edge channel) widths are calculated imposing electrostatic equilibrium and by solving the related Poisson equation utilizing holomorphic functions and analytical continuity. Then the width of the $k^{\rm th}$ IS (where, $k$ is an even integer corresponding to $\nu$) is given by, 
 \begin{equation}\label{ak}
     a_{k}=\left(\frac{2\epsilon\Delta E_{k}}{\pi^{2}e^{2}\frac{dn(x)}{dx}|_{x_{k}}}\right)^{\frac{1}{2}},
  \end{equation}
where $\epsilon$ is the dielectric constant, $n(x)$ is the local electron density and $\Delta E_{k}=\hbar\omega_{c}$ is the single particle energy gap between two successive Landau levels and $\omega_{c}=\frac{eB}{m}$ is the cyclotron frequency. The local electron density is obtained as,
\begin{equation}\label{nx}
   n(x)=\left(\frac{x-l_{d}}{x+l_{d}}\right)^{\frac{1}{2}}n_{0}, x>l_{d},
  \end{equation}
where $l_{d}$ is the depletion length and  $n_{0}$  is the bulk electron density, far from the edges. In the next section we will utilize and obtain the curvature modified given equations to calculate the properties of ISs residing on a curved structure. We assume that the electrostatics is unaffected by the curvature, which is justified by experiments~\cite{Friedland20081087}. The important difference between the planar and the curved structure will be the spatially varying $B$ field along the current direction, that affects the energy gap locally hence the formation of ISs.

\textbf{Curved interferometer}

A non-planar 2DES (GaAs/ AlGaAs) can be created at self-rolled heterostructures~\cite{Friedland20081087}. The multi-layered heterostructure is rolled up into a tube with a constant radius of curvature $r$ during selective etching to fabricate a cylindrical surface that comprises a 2DES \cite{Prinz2000828,mendach:212113, Vorob'ev2004171}. To investigate the motion of an electron on a curved structure, the low temperature mean free path of an electron $l_{s}$ should be comparable to $r$\cite{PhysRevB.79.125320}. 

In the presence of an external $B$ field directed along the $z$ axis, the spinless electrons experience a spatially varying field, since Landau quantization depends only on the perpendicular field component $B_{\perp}(=B\cos(\theta))$. Hence, on a curved 2DES (C2DES) the widths of the ISs will vary depending on the location along the current direction and Eq.~\ref{ak} is modified accordingly yielding,
\begin{equation}\label{ak1}
 a_{k}(\theta)=\sqrt{\frac{16a_{B}l_{d}(\theta)k}{\pi}}\frac{\nu_{0}(\theta)}{\nu_{0}^{2}(\theta)-k^{2}},
\end{equation}
where  $a_{B}$ is the effective Bohr radius ($a_{B}^{GaAs} \sim$ 9.8 nm) and $\nu_{0}(\theta)=2\pi n_{0}\ell_{B}^{2}(\theta)$ is the filling factor at the bulk, which depends on the location along the curved Hall bar. In addition, to form a closed loop, we impose two QPCs on the device which are defined on the top of the C2DES by split-gates. Fig.~\ref{fig1} depicts the interferometer defined on a curved structure. Recall that the width and the centre of the channel depend strongly on the ``constant" depletion length along the Hall bar at a planar structure, which is controlled by gate voltage as $l_d=(V_g\epsilon)/(4\pi^2n_0e)$. At a curved structure one can assume that this depletion length also depends on the location of the QPC's. Thus we define QPCs on a C2DES by imposing position dependent (in fact angle) depletion length assuming commonly utilised Gaussian form as, 
  \begin{equation}\label{ld}
   l_{d}(\theta)=\tilde{V_{g}} \left[e^{-\left(\frac{(\theta-\theta_{\rm L})^{2}}{2\alpha^{2}}\right)}+e^{-\left(\frac{(\theta-\theta_{\rm R})^{2}}{2\alpha^{2}}\right)}
   \right],
  \end{equation}
where $\tilde{V_{g}}$ is the normalized gate voltage, $\alpha$ determines the width of the QPC and $\theta_{\rm L,R}$ fix the centres of the QPCs, at left and right sides of the quantum dot (QD), respectively.

\begin{figure}
     \scalebox{0.3}{\includegraphics{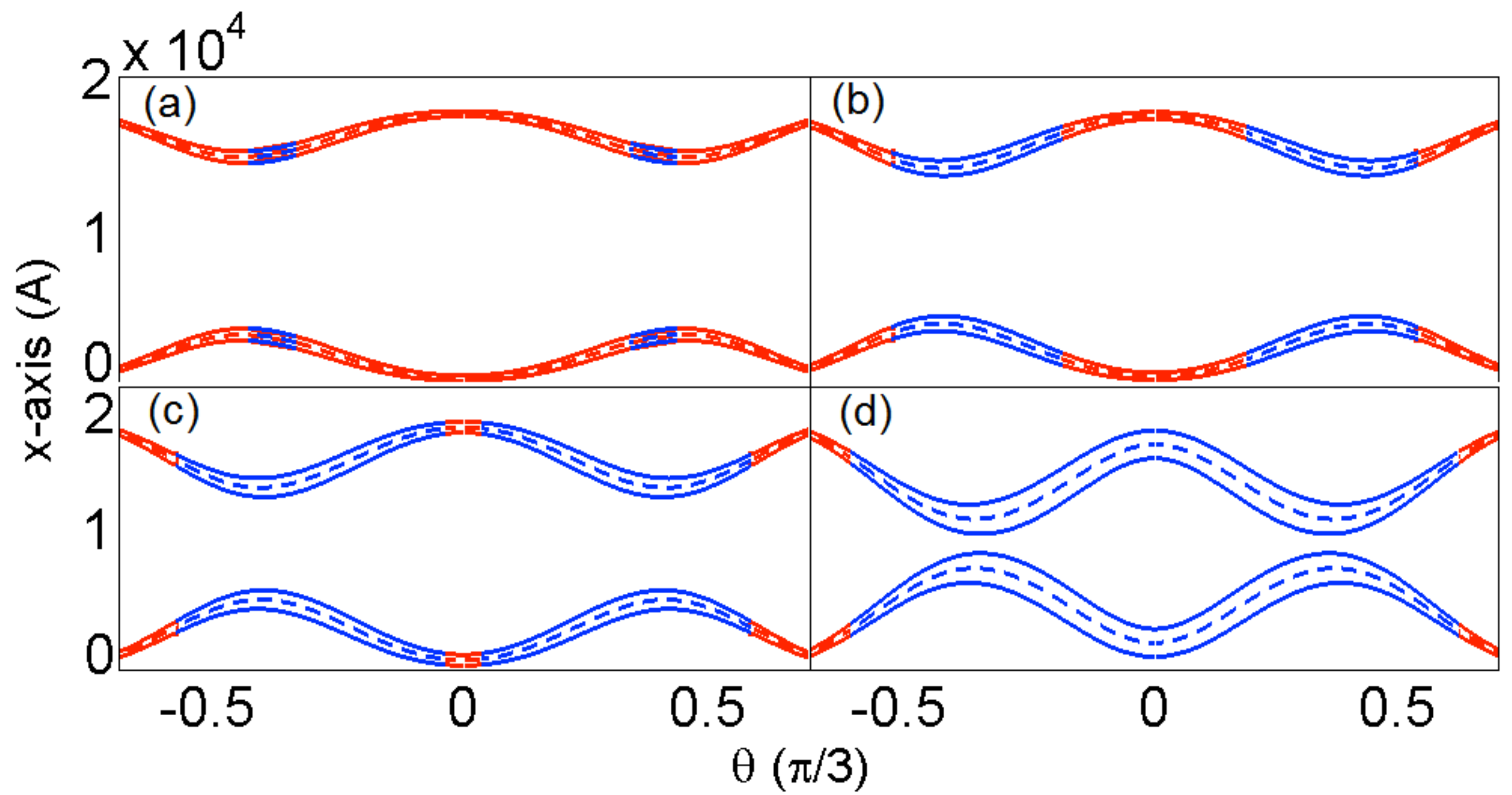}}
    \caption{The spatial distribution and the widths of the ISs calculated at characteristic field values, (a) 1.7 T, (b)1.8 T, (c)1.9 T and (d) 2 T. The device is taken to be 2 $\mu$m wide and 10 $\mu$m long, assuming typical experimental bulk density, $n_0\sim 1\times10^{15}$ m$^{-2}$. The QPCs reside at $\theta_{L,R}=\pm\pi/6$, $\alpha=\pi/12$ and the depletion length at the centre of QPC's ($\theta=\theta_{L,R}$) is set $l_d^C=500$ {\AA}.}\label{fig2}
\vspace{-.6 cm}
\end{figure}

In Fig.~\ref{fig2}, we show the spatial distributions of $\nu=2$ incompressible strips calculated at characteristic field strengths. The broken lines denote 1D edge channels, whereas the ISs are formed between the solid lines. The dark coloured regions (blue) point a well developed strip (i.e $a_2>\lambda_F$). Evanescent strips ($\ell_B<a_2<\lambda_F$) are depicted by slightly light colour (red). Recalling our previous discussion on the observation conditions of the interference pattern, we see that it is only possible to get conductance oscillations in the close vicinity of $B=1.9$ T, where the strips are evanescent and are sufficiently close to each other. 

Next, we take into account Zeeman spin-splitting to determine the widths of the odd-integer ISs, i.e considering electrons with spin. The energy gap $\Delta E_{k}$ has different values for even and odd filling factors, $[\Delta E_{k}]_{even}=\hbar\frac{eB_{\perp}}{m^{*}}-g^{*}\mu_{B}B$ and $[\Delta E_{k}]_{odd}=g^{*}\mu^*_{B}B$, where $\mu^*_{B}$ and $g^{*}$ are the effective Bohr magneton and the Land\'{e} $g$ factor, respectively. Thus the widths of the ISs are determined by,
\begin{equation}\label{akeven}
    [a_{k}(\theta)]_{even}=\sqrt{\frac{8a_{B}l_{d}(\theta)}{\pi}}\frac{\nu_{0}(\theta)\sqrt{k}}{\nu_{0}^{2}(\theta)-k^{2}}\left[2-\frac{g^{*}}{\cos\theta}\right],
  \end{equation}
and
  \begin{equation}\label{akodd}
    [a_{k}(\theta)]_{odd}=\sqrt{\frac{8a_{B}l_{d}(\theta)}{\pi}}\frac{\nu_{0}(\theta)\sqrt{k}}{\nu_{0}^{2}(\theta)-k^{2}}\sqrt{\frac{g^{*}}{\cos\theta}}
  \end{equation}
Note that, the Zeeman energy depends on total $B$ field, whereas the Landau energy depends on both the total and the perpendicular field.   
  
\begin{figure}
\centering
\includegraphics[width=1\columnwidth]{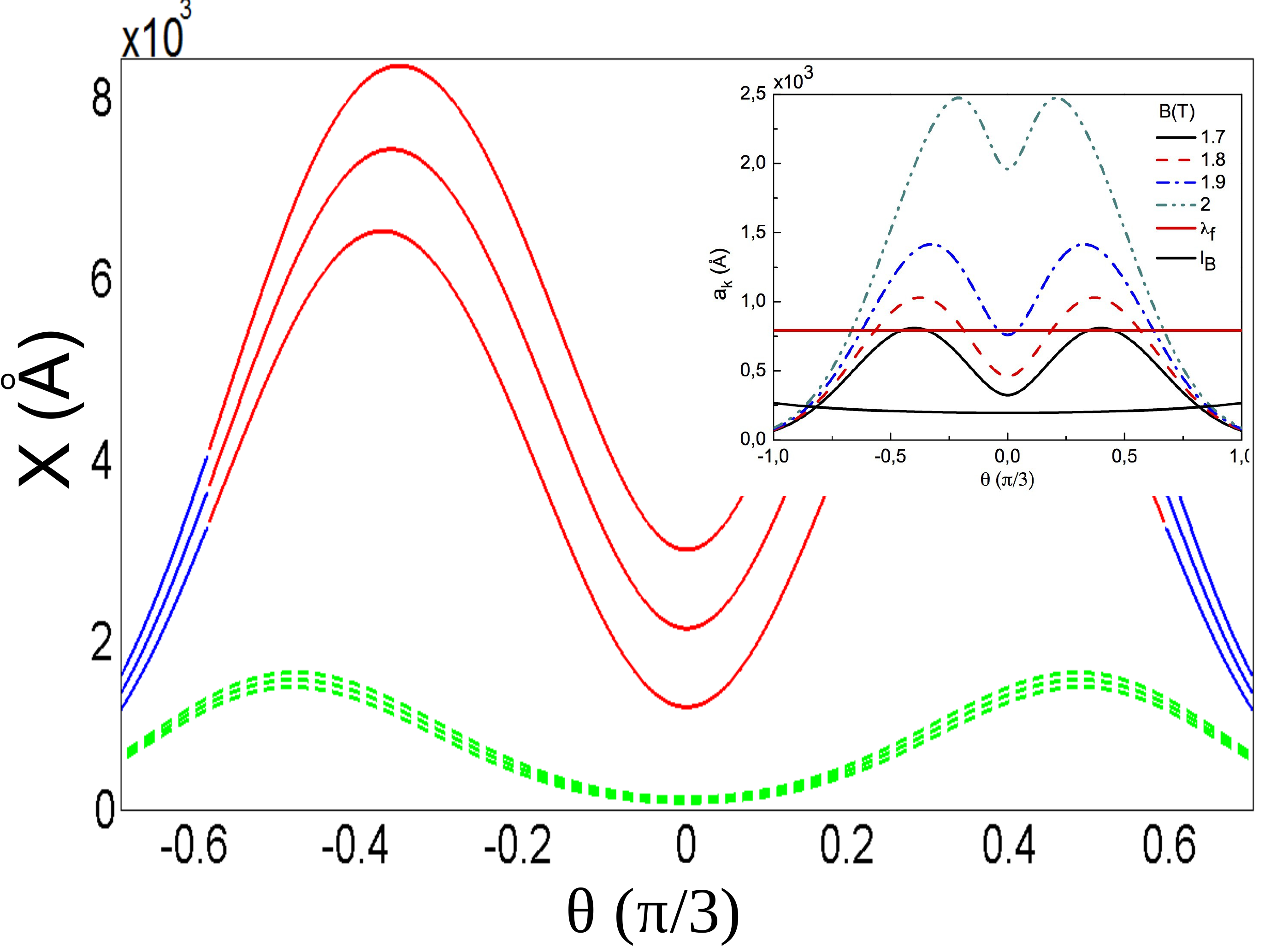}
\caption{ \label{fig3} (a) The spatial distribution of $\nu=1$ (lower, green) and 2 (upper, red-blue) ISs along the device, central lines indicate edge-states. Inset: Widths of ISs along the device considering $\nu=2$.}
\end{figure}

Fig.~\ref{fig3}, plots the distribution of ISs of $\nu=1$ and 2 along the device. We observe that, the outer strip assuming $\nu=1$ is barely affected by the curvature and is evanescent. Meanwhile the inner strip with $\nu=2$, satisfies interference conditions. At the proposed interference experiments performed on curved structure, the different open loop of odd $\nu$ and closed loop of even $\nu$ would be detected by different oscillation periods in $\Delta B$. To be explicit, if the period of even $\nu$ is not twice the odd $\nu$, i.e. $\Delta B_{\nu=2}\neq2.\Delta B_{\nu=1}$, then this will impose that the origin of the conduction oscillations stems from the AB phase of the particles, but not due to charging effects. Since, charging effects would not depend on the area without a top gate, at small samples. In the inset of Fig.~\ref{fig3}, we show the width of ISs at different field strengths along the device for $\nu=2$. A strong field dependence of $\nu=2$ strip is observed, where at $B=1.7$ we do not expect an interference, since the strips are in the diffusive regime. For larger field values (1.8 T and 1.9 T), strips near the QPCs become evanescent and incompressible within the QD, hence interference is expected. At $B=2.0$ T, the strip is incompressible at the constrictions and no partitioning can take place, therefore no conduction oscillations can be observed.       

\begin{figure}
\centering
\includegraphics[width=1\columnwidth]{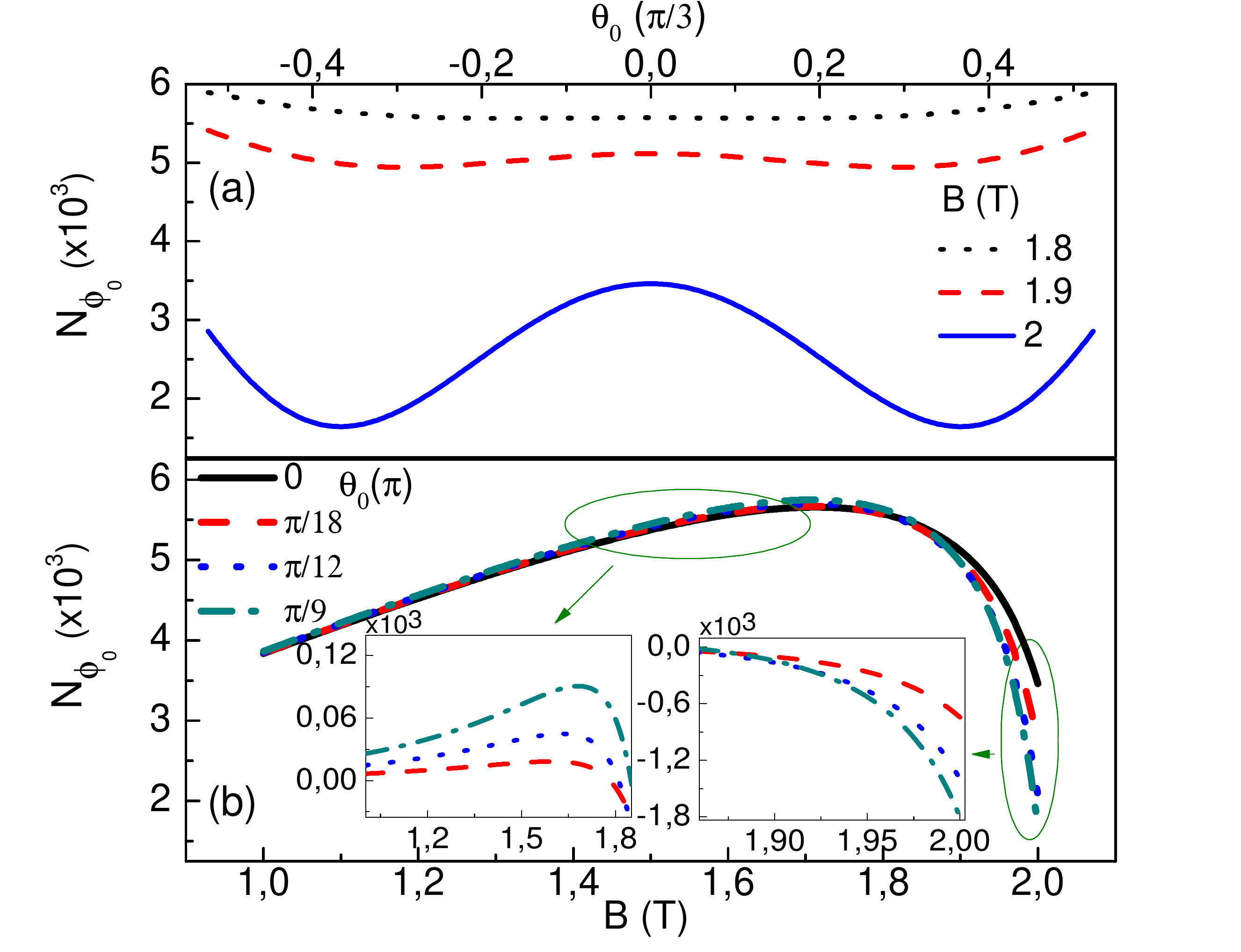}
\caption{ \label{fig4} (a) The number of flux quanta calculated as a function of rotation angle $\theta_0$ at three field strengths. (b) The same quantity as a function of field for different values of fixed $\theta_0$.}\vspace{-.5cm}
\end{figure}

Next, we focus on the variation of $N_{\Phi_0}$ as a function of field and rotation angle $\theta_0$. In Fig.\ref{fig4}a, we show $N_{\Phi_0}$ enclosed in the central part of the dot by $\nu=2$ ISs considering three field strengths for which we expect to observe interference. As we rotate the sample with respect to $z$ axis, $N_{\Phi_0}$ first decreases in a non-linear manner and then increases at an angle $\theta\sim\pi/9$. While rotating the interferometer, first the area of the loop decreases till two separate loops are formed, then the loop residing at either side starts to enlarge, hence, $N_{\Phi_0}$ increases. The above described situation is most pronounced at 2T, however, for the lowest $B$ the variation in $N_{\Phi_0}$ is mostly monotonous. Once again, we predict that if the observed conduction oscillations are due to charging effects the period should not be affected when rotating the sample, while without a top gate the capacitance should not depend on the area of the device. Fig.\ref{fig4}b, depicts the evolution of $N_{\Phi_0}$ as a function of field for $\nu=2$ considering four different central angle $\theta_C$. Recall that, while changing the field the positions of the ISs also change, hence, the area of the closed loop changes. The area of the loop remains almost constant at low fields, therefore, increasing the field only results in a linear increase of the flux number. However, at higher fields the area starts to shrink. As a consequence, although the field increases $N_{\Phi_0}$ remains constant for a narrow $B$ interval and even decreases by increasing field. Constant $N_{\Phi_0}$ in $B$ implies oscillations to be smeared. In addition, by rotating the sample one observes that the linear behaviour is altered and the number of flux changes differently at different rotation angles. The insets of Fig.\ref{fig4}b, plots $[N_{\Phi_0}(\theta_{0}=0)-N_{\Phi_0}(\theta_{0}=\theta_{C})]$ and we see that under rotations $N_{\Phi_0}$ is larger for larger angles at the linear regime and is smaller at high field regime ($B>1.9$ T).     
As a final remark, we would like to discuss the limitations of our model: First of all, the sample width is considered to be much larger than the magnetic length, since the Thomas-Fermi Approximation, namely potential varies slowly,  is only valid at this limit. Second, sample length should be smaller than the mean free path and coherence length so that one can observe interference effects. Also note that, we utilised the electrostatic approximation which is well defined on a planar structure which might be questioned to be used at a curved structure. However, this assumption is in line with slowly varying potential both in lateral and current direction, hence is justified.

\textbf{Conclusion}

We investigated the edge-state distribution at a Fabry-P\'erot interferometer defined on a curved heterostructure. We observed that, the properties of the interference channels strongly depend on the curvature, both in widths and more importantly by the area enclosed by them. It is observed that, the loop area varies with field strength linearly at low fields and, shrinks at the high fields. As a direct consequence the number of flux quanta penetrating the loop varies, also as a function of the rotation angle. The additional tuneable parameter, the rotation angle, brings a novel technique to probe the origin of conduction oscillations observed. Since, by rotating the sample one can change the area while keeping the number of particles within the dot constant. Hence, in the proposed experiments one can distinguish between Coulomb Blockade (due to charging) regime and the Aharonov-Bohm regime (due to phase). In addition, due to different $B$ dependency of Zeeman and Landau energies one can observe different areal dependencies of the oscillation periods by investigating the difference in odd-even filling factors.

T\"UB\.ITAK is acknowledged for financial support under grants 112T148 and 211T264







\end{document}